\documentclass[10pt,a4paper]{article}
\usepackage[utf8]{inputenc}
\usepackage{amsmath}
\usepackage{amsfonts}
\usepackage{amssymb}
\usepackage{hyperref}
\usepackage{graphicx}
\usepackage{caption}
\usepackage{subcaption}
\usepackage{float}
\usepackage{cite}

\usepackage[left=2.2cm,right=2.2cm,top=2.5cm,bottom=2.5cm]{geometry}
\usepackage[affil-it]{authblk}

\title{Quantum gravity effects in statistical mechanics with modified dispersion relation
}
\author{Shovon Biswas$^{a}$}
\author{Mir Mehedi Faruk$^{b,c}$\thanks{muturza3.1416@gmail.com,  mir.faruk@mail.mcgill.ca}}
\affil{\small{Department of Electrical and Electronic Engineering, Bangladesh University of Engineering and Technology, Dhaka 1000, Bangladesh $^a$}}
\affil{Bose Centre for Advanced Study and Research in Natural Science, University of Dhaka, Bangladesh. $^b $}
\affil{Physics Department, McGill University Montreal, QC H3A 2T8, Canada $^c$}
 
\begin{document}
\maketitle
\begin{abstract}
    Planck scale inspired theories  which are also often  accompanied with maximum energy and/or momentum scale predict deformed dispersion relations compared to ordinary special relativity and quantum mechanics. 
    In this paper 
    we resort to the methods of statistical mechanics in order to
determine the effects of a deformed dispersion relation along with 
upper bound in partition function as maximum energy and/or momentum scale can 
have on the thermodynamics
of photon gas. We have analyzed two distinct quantum gravity models in this paper.
\end{abstract}

\section{Introduction}
Different quantum gravity approaches such as string theory, loop quantum gravity, noncommutative geometry, doubly special relativity (DSR), generalized uncertainty principle (GUP) suggest the existence of a minimal length scale of the order of Planck length $l_p=\sqrt{\frac{G\hbar}{c^3}}$. This length and its inverse, the Planck energy $E_p$, mark thresholds beyond which the old description of spacetime breaks down and qualitatively new phenomena are expected to appear\cite{mag1}. 
DSR is a  very well known approach to deal with minimal length situations. DSR type  theories  predict a modified dispersion relation compared to usual dispersion relation of special relativity. But one should notice that DSR 
can be formulated in different approaches.
One of the approaches \cite{zhang} which maintain 
varying speed of light (VSL) scenario
can be entitled with maximum momenta or maximum energy depending on the value of 
parameter which describes the minimal length scale of the theory (We will refer this as model A in this manuscript).. Another possible DSR scenario\cite{daso,mmfc} in which speed of 
light  is constant is also heavily studied with the help of
nonlinear realizations of the Lorentz group (This is referred  as model 1 in the rest of the paper). Point to note these different models predict different dispersion relations and distinct physical phenomena. The implications of non-zero minimal length has been considered recently in different contexts that include harmonic oscillator
\cite{osci1,osci2,osci3,osci4,osci5,osci6}, hydrogen atom \cite{atom1,atom2,atom3,atom4}, gravitational quantum well \cite{qwell},
the Casimir effect \cite{casi1,casi2}, particles scattering \cite{sca1,sca2}, relativistic thermodynamics of ideal fluid \cite{das1}, blackbody radiation \cite{bb1} etc. It has been reported\cite{dop} and well established that
a consistent marriage of ideas of
quantum mechanics and gravitation needs a noncommutative
description of spacetime to avoid the paradoxical
condition of creating of a black hole for an event that is sufficiently localized in spacetime. 
As DSR admits a noncommutative spacetime,
DSR theory fits the
criterion for being a quantum gravity framework. 
 \\\\
Invariant energy scale, i.e. Planck scale  modified dispersion relations are usually identified with maximum energy or maximum momenta or both\cite{mag1,mag2,zhang}.
It is interesting to note that, dispersion modifications can lead to many further
predictions which can be tested in a series of experiments such as Gamma ray burst observation\cite{am1} or inflationary cosmology\cite{inflation}. In this spirit lots of studies are done on statistical thermodynamics of quantum gases\cite{daso,mmfc,nozari,farag}. One has to be very careful while calculating the partition function of the quantum gases within these models as different models change the structure of partition function. For instance, in case of of massless boson, model A changes both the integration measure and upper limit of the partition function\cite{zhang} while in model 1 only the upper limit is changed\cite{daso,mmfc,cam}. Different errors have been noticed in the several manuscripts while evaluating the partition function for quantum gases. For example, both in reference \cite{daso} and \cite{nozari} Maxwell Boltzmann distribution functions are used to calculate  the thermodynamics of photon gas.  Ref\cite{daso} is based on model 1. One of us (MMF) had
recently took an attempt\cite{mmfc} to evaluate the partition function of photon gas  with correct distribution function, i.e. Bose-Einstein distribution function just as Ref.\cite{zhang}. But while calculating the partition function in  previous manuscript\cite{mmfc} an error has been noticed, which will be corrected in this paper (see footnote 2 for details).
The deformed dispersion relation predicted by model 1 can be written as\cite{mag1,daso},
\footnote{we have adopted $\hbar=c=k_B=1$}
\begin{equation}
   E^2=p^2+ m^2(1-\frac{E}{E_p})^2 .
\end{equation}
Here, $E$ and $p$ are the energy and the magnitude of the three momentum
of the particle, respectively, while $m$ is the mass
of the particle and $E_p$ is the Planck energy. Now for massless particles the dispersion reduces to $E=p$. That is why an upper energy bound in energy is equivalent to upper energy bound in momentum in this case. However, in model A the dispersion relation for massless particles\cite{zhang} takes the form, 
\begin{equation}
    p=\frac{E}{1+\lambda E}.
\end{equation}
where $ \mid \lambda \mid\propto  \frac{1}{E_P } $, so $\lambda$ can be both positive and negative.
Now when $\lambda > 0$, Eq. (2) implies a maximum momentum $p_{max}=\lambda^{-1}$ ,
which remains invariant under deformed transformation laws.
In contrast, $\lambda < 0$
corresponds to an energy upper bound for photons. So, in case of model A we find  an 
upper bound in either momentum or in energy, whereas in model 1 we see an upper bound both in energy and momentum. In this manuscript we will solve the partition function for photon gas with model 1.
\\ \\ 
Now,
another well studied modified dispersion relation based on phenomenological study is\cite{cam,am2},
\begin{equation}
    E^2=p^2 [1-\alpha (\frac{E}{E_P})^n] +m^2.
\end{equation}
Here, $E_P$ is the Planck energy and  $\alpha$ is a coefficient of order 1, whose precise value depends upon the considered
quantum-gravity model, while $n$, the lowest power in Planck’s length leading to
a non-vanishing contribution, is also model dependent
quantity. This type of dispersion relation is observed in loop quantum gravity\cite{am2}.
In ref.\cite{am2, am3,am4,gil} it has been argued $n$ can be chosen as $n=1$ or $n=2$. See ref. \cite{am2} to understand the different physical scenarios with $n=1$ and $n=2$. Now
choosing $m=0$ and 
following the spirit of Zhang et. al\cite{zhang}  
 we find out from eq. (3) that,  $E_P>0$ indicates an upper bound in energy $E_{max}=E_P$. A bound in momentum can be obtained with $E_P<0$, which is unphysical, so we discard the scenario. 
We  must note that Camacho and Marcias \cite{cam} have already visited thermodynamics of photon gas within this model. But they did not take into account the upper bound of maximum energy while calculating the partition function.
In this paper, we will also
visit the thermodynamics of photon gas with this  deformed dispersion relation, but with maximum energy bound as argued above. We will refer this model as model 2 in this paper.
\section{Thermodynamics of Photon gas with deformed dispersion relation in model 1}
\subsection{Partition Function} 
The expression of partition function in usual statistical mechanics for massless bosons  in grand canonical ensemble can be obtained\cite{path},
\begin{equation}
    \ln Z= -\sum_i \ln (1- e^{-\beta E_i}).
\end{equation}
Here, $\beta=\frac{1}{T}$. 
Now changing the sum to integral we find out, 
\begin{equation}
            \ln Z=-\frac{1}{(2\pi)^3}\int \int\;  d^3r d^3p \ln (1- e^{-\beta E}).
\end{equation}
In ref.\cite{mag1},  Magueijo and Smolin proposed a DSR model where 
the dispersion relation for a massive particle takes the form,
\begin{equation}
    E^2=p^2+m^2\left(1-\frac{E}{E_p}\right)^2.
\end{equation}
For massless photon gas equation (6) takes the usual form $E=p$. Take a look in ref.\cite{daso} to see how the phase space volume remains unaltered  in this model. Now, due to the presence of upper bound in energy, $E_P$ in this model, the partition function  becomes,
\begin{equation}
            \ln Z= -\frac{4\pi V}{(2\pi)^3}\int_0 ^{E_P} E^2 \ln (1- e^{-\beta E})dE.
\end{equation}
Here, $V$ is the total volume of the system. Eq (7) can be written as,
\begin{equation}
    \ln Z=\frac{4\pi V}{(2\pi)^3}\frac{T^3}{3}f_4.
\end{equation}
where the function $f_n$ is defined in Appendix.\footnote{In order to calculate eq (7) one can choose $x=\beta E$. As a result, the upper limit of integration should also be changed accordingly. Although correct distribution function, i.e.
Bose-Einstein distribution was taken instead of  Maxwell-Boltzmann distribution  in our study\cite{mmfc},  the change of the upper limit of integration with respect to this variable shift was not mistakenly taken into account. Once this variable shift is also taken into account in the upper bound of the integration, one finds out the upper bound of the integration contains temperature dependent term. which was  completely absent in ref. \cite{mmfc}.}
\subsection{Thermodynamic Quantities}
Thermodynamic quantities can be derived easily from the partition function following the simple and effective rules of statistical mechanics\cite{path}. For example, the internal energy is related to partition function by
\begin{eqnarray}
    U&=&T^2\left(\frac{\partial \ln Z}{\partial T}\right)_V \nonumber\\
    &=&\frac{4\pi V}{(2\pi)^3}T^4\left[f_4+\frac{T}{3}f'_4\right].    
\end{eqnarray}
The pressure can be found from the relation\
\begin{eqnarray}
    PV&=&T\ln Z\\
    &=&\frac{4\pi}{(2\pi)^3}\frac{T^4}{3}f_4.
\end{eqnarray}
We see that the well known relationship between pressure and energy $P=\frac{U}{3V}$ is not valid anymore as obtained in usual statistical mechanics.
Another important thermodynamic quantity entropy $S$, 
\begin{equation}
    S=-(\frac{\partial F}{\partial T})_V=\frac{4\pi V}{(2\pi)^3}T^3\left[\frac{4}{3}f_4+\frac{T}{3}f'_4\right].
\end{equation}
where, $F=-Tln Z$. As $T\rightarrow 0$, we notice $S\rightarrow 0$ indicating that Nernst's postulate is valid.
Finally specific heat can be obtained from the relation
\begin{eqnarray}
C_V  &&  =\left(\frac{\partial U}{\partial T}\right)_V\nonumber\\
      &&  =\frac{4\pi V}{(2\pi)^3}T^4\left[ \frac{4f_4}{T}+\frac{8}{3}f'_4+\frac{T}{3}f^{''}_4\right].
\end{eqnarray}
All the thermodynamic quantities reduce to usual results of statistical mechanics in the limit $E_P\rightarrow\infty$\cite{path}.
\section{Thermodynamics of Photon gas with deformed dispersion relation in model 2}
\subsection{The partition function}
Now let us focus on the model based on phenomenological study, first discussed in detail by  Amelino-Camelia\cite{am1,am2}. Later thermodynamics of photon gas were studied 
within this model by Camacho and Marcias\cite{cam}. As discussed in introduction Camacho and Marcias did not take into account the upper energy bound that exists in this model, while evaluating the partition function. In this section we evaluate the partition function in presence of the upper  bound.\\\\
The modified dispersion relation of
massless bosons in this model
\begin{eqnarray}
    &&E^2=p^2\left[1-\alpha(E/E_P)^n\right] \nonumber\\
   \Rightarrow && \frac{1}{p^2}=\frac{1-\alpha(\frac{E}{E_P})^n}{E^2}.
\end{eqnarray}
Now it is clear from above equation with $p\rightarrow \infty$, $E$ tends to $\frac{1}{\alpha^{1/n}}E_P$, i.e. the maximum value of $E$, considering  $E_P>0$. As $\alpha\sim 1$\cite{am1,am2}, we find out $E_{max}=E_P $. \\\\
\\
The partition function can be equivalently written as,
\begin{equation}
            \ln Z= -\int dE  \rho(E)  \ln (1- e^{-\beta E}).
\end{equation}\\
where $\rho(E)$ is the density of states\cite{cam},
\begin{eqnarray}
    \rho(E)=\frac{4\pi V}{(2\pi)^3}E^2[1+(n+3/2)(E/E_p)^n]
\end{eqnarray}
Based on above observation we can write the partition function, 
\begin{equation}
    \ln Z=-\frac{4\pi V}{(2\pi)^3}\int_{0}^{E_p}E^2\left[1+(n+3/2)(E/E_p)^n \right] \ln [1-e^{-E/T}] dE.
\end{equation}
Equation (17) can be rewritten as
\begin{equation}
\begin{split}
\ln Z&=\frac{4\pi V}{(2\pi)^3}\left[ I_4+\frac{(n+3/2)}{E_p^n}I_{n+4}\right]\\
    &= \frac{4\pi VT^3}{(2\pi)^3}\left[ \frac{f_4}{3}+\frac{(n+3/2)}{n+3}\left(\frac{T}{T_p}\right)^nf_{n+4}\right].
\end{split}
\end{equation}

\subsection{Thermodynamic quantities}
Just like section 2.2, the thermodynamic quantities can be derived from the partition function.\\ \\
Internal energy,
\begin{equation}
    U=T^2\left(\frac{\partial \ln Z}{\partial T}\right)_V=\frac{4\pi V}{(2\pi)^3}T^4\left[ f_4+\frac{T}{3}f'_4+(n+3/2)\left(\frac{T}{T_p}\right)^n\left\{ f_{n+4}+T\frac{f'_{n+4}}{n+3}\right\}\right].
\end{equation}
Pressure, 
\begin{equation}
    P=\frac{4\pi T^3}{(2\pi)^3}\left[ \frac{f_4}{3}+\frac{(n+3/2)}{n+3}\left(\frac{T}{T_p}\right)^nf_{n+4}\right].
\end{equation}
Entropy, 
\begin{equation}
    S=\frac{4\pi V}{(2\pi)^3}T^3\left[\frac{4}{3}f_4+\frac{T}{3}f'_4+\frac{(n+3/2)}{n+3}\left(\frac{T}{T_p}\right)^n\left\{(n+4)f_{n+4}+T{f'_{n+4}}\right\}\right].
\end{equation}
Specific heat, 
\begin{eqnarray}
  C_V && =\left(\frac{\partial U}{\partial T}\right)_V\nonumber\\
    &&=\frac{4\pi V}{(2\pi)^3}T^4\Bigg[ \frac{4f_4}{T}+\frac{8}{3}f'_4+\frac{T}{3}f^{''}_4 + (n+3/2)\left(\frac{T}{T_p}\right)^n\bigg\{ \frac{(n+4)f_{n+4}}{T}+\nonumber\\
    &&\frac{2(n+4)}{n+3}f'_{n+4}+\frac{T}{n+3}f^{''}_{n+4}
    \bigg\}\Bigg]
\end{eqnarray}

The thermodynamic quantities of this model also coincides
with the results of usual statistical mechanics in the limit $E_P \rightarrow \infty$. The presence of upper bound in the partition function drastically changes the final outcoe of this model. Several other observations will be noted in the next section.

\section{Discussion and Conclusion}
We are now in a position to present the results, obtained in previous two sections. The results have been obtained using $Mathematica$.\\ \\
In figure 1 we have plotted the pressure obtained from model 1 and model 2, with $n=1$ and $n=2$, and compared with the usual results of special relativity (SR). In the study of model 1 (DSR), we have noticed
that the values of pressure exactly coincides with SR results in  low temperature region but shows deviation from usual result at high temperature limit.  At any given temperature, the  thermodynamic quantities in DSR model pick smaller value compared to usual result in that temperature. This is expected  in the DSR model,  as the dispersion relation is unchanged compared to SR and due to the presence of upper energy scale the total number of microstates becomes less than SR.
Although this result has been reported before in ref. \cite{mmfc} we should note that,
 due to the correction (see footnote 2) we have presented in the current paper in evaluating the partition function for DSR model,  the resulted   thermodynamic quantities take lesser value compared to the previous study\cite{mmfc}.   \\\\
In the previous study with model 2, Camacho and Marcias have predicted that, the pressure as well as other thermodynamic quantities are always greater than the SR results.
This scenario is drastically changed in our study as we have taken into account the existence of 
upper energy bound $E_P$.
At low temperature the predicted pressure of this model coincide with the with usual result as it should be but after a certain temperature, it starts taking larger value compared to usual results.
Near, $T=0.5T_P$
the model 2 result intersects with SR result and afterwards the predicted outcome
take lesser value compared to usual result of  SR. 
This is due to the cut off we considered in our calculation. Such property is also noticed in recent Lorentz violating study\cite{mmf.kazi} with upper cutoff. 
This characteristic is also noticed in case of other thermodynamic quantities of model 2 (see figure $1b$ and $1c$).
This type of unique characteristic  is not noticed when an  upper bound in the partition function\cite{cam} is not present.\\\\
The different thermodynamic quantities  for ideal  photon gas in SR theory maintain a specific relation, such as-
pressure-energy relation ($(PV=3U)$), internal energy-free energy relation $(U=-3F)$, specific heat-entropy relation $(3S=C_v)$\cite{path}. None of them remain valid for any of these  models with upper energy bound.  In the previous study with model 1\cite{mmfc}, it was reported that such relations are maintained but
it was due to the error we pointed out in footnote $2$. Once we have taken care of the error the relations are not maintained anymore.
\begin{figure}[H]
\begin{subfigure}{0.5\textwidth}
  \centering
  \includegraphics[width=0.98\linewidth]{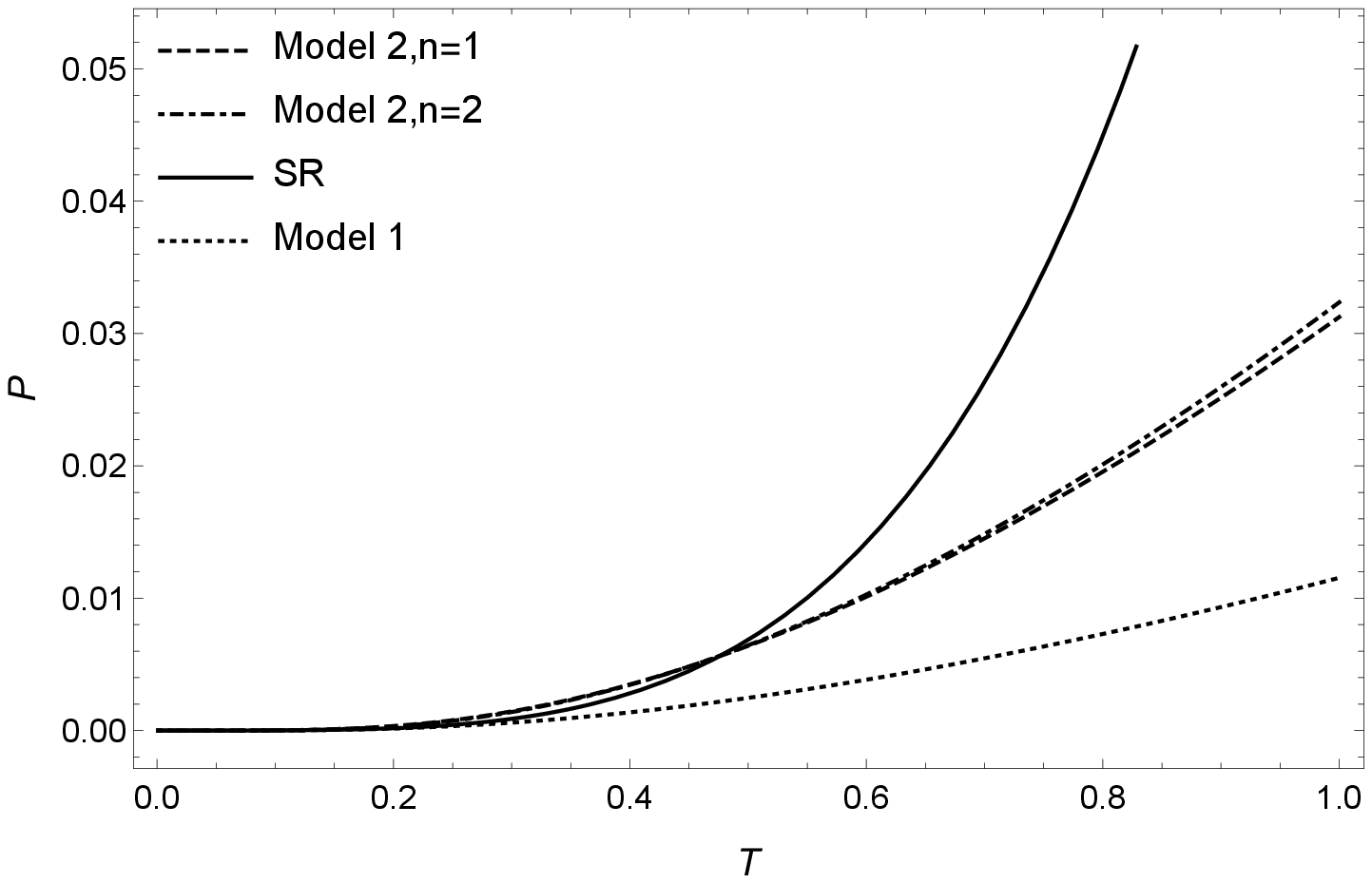}
  \caption{Pressure $P$, versus temperature}
  \label{fig:sub1}
\end{subfigure}%
\begin{subfigure}{.5\textwidth}
  \centering
  \includegraphics[width=0.98\linewidth]{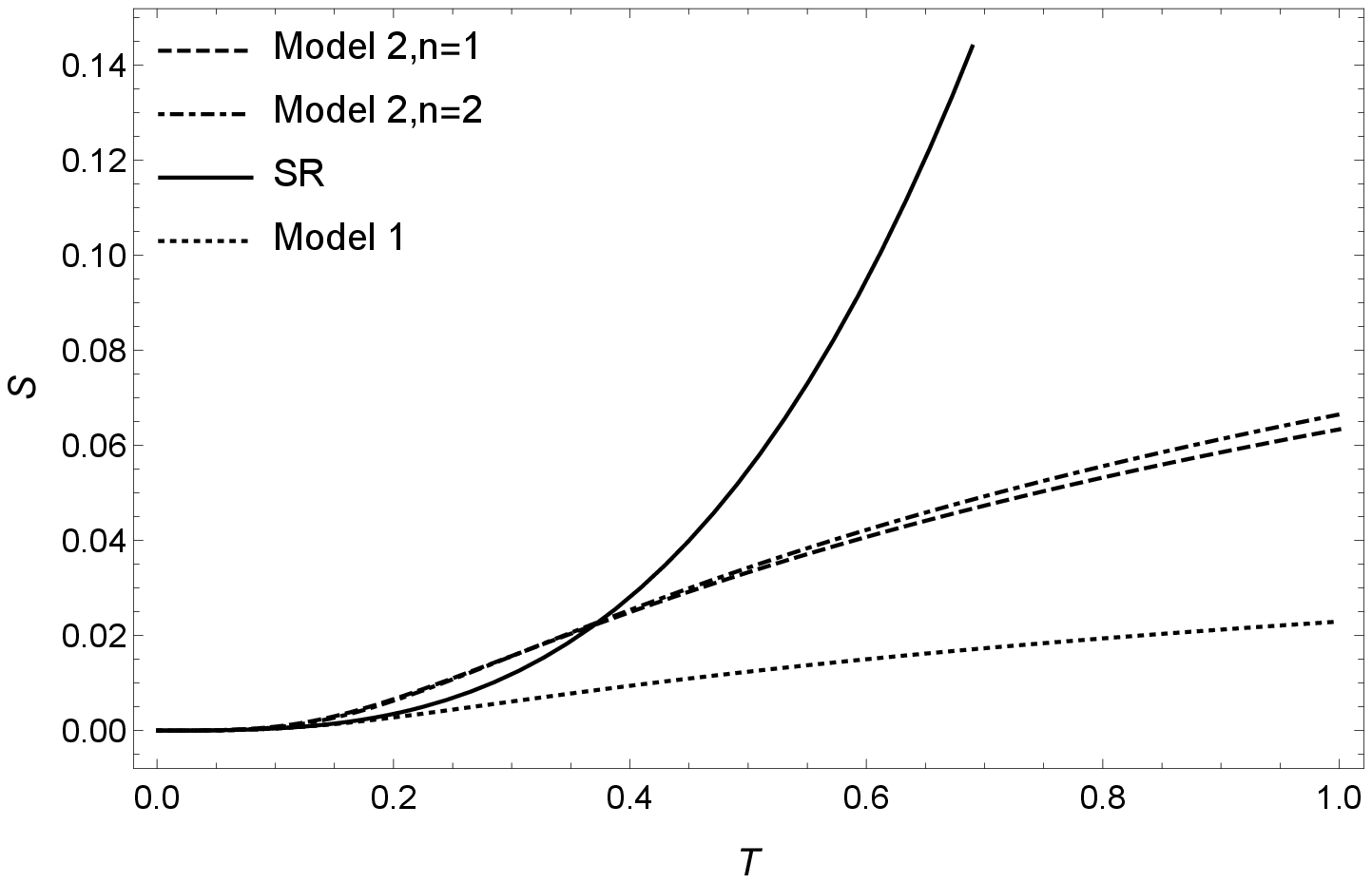}
  \caption{ Entropy $S$, versus temperature }
  \label{fig:sub2}
\end{subfigure}
\begin{subfigure}{.5\textwidth}
  \centering
  \includegraphics[width=0.98\linewidth]{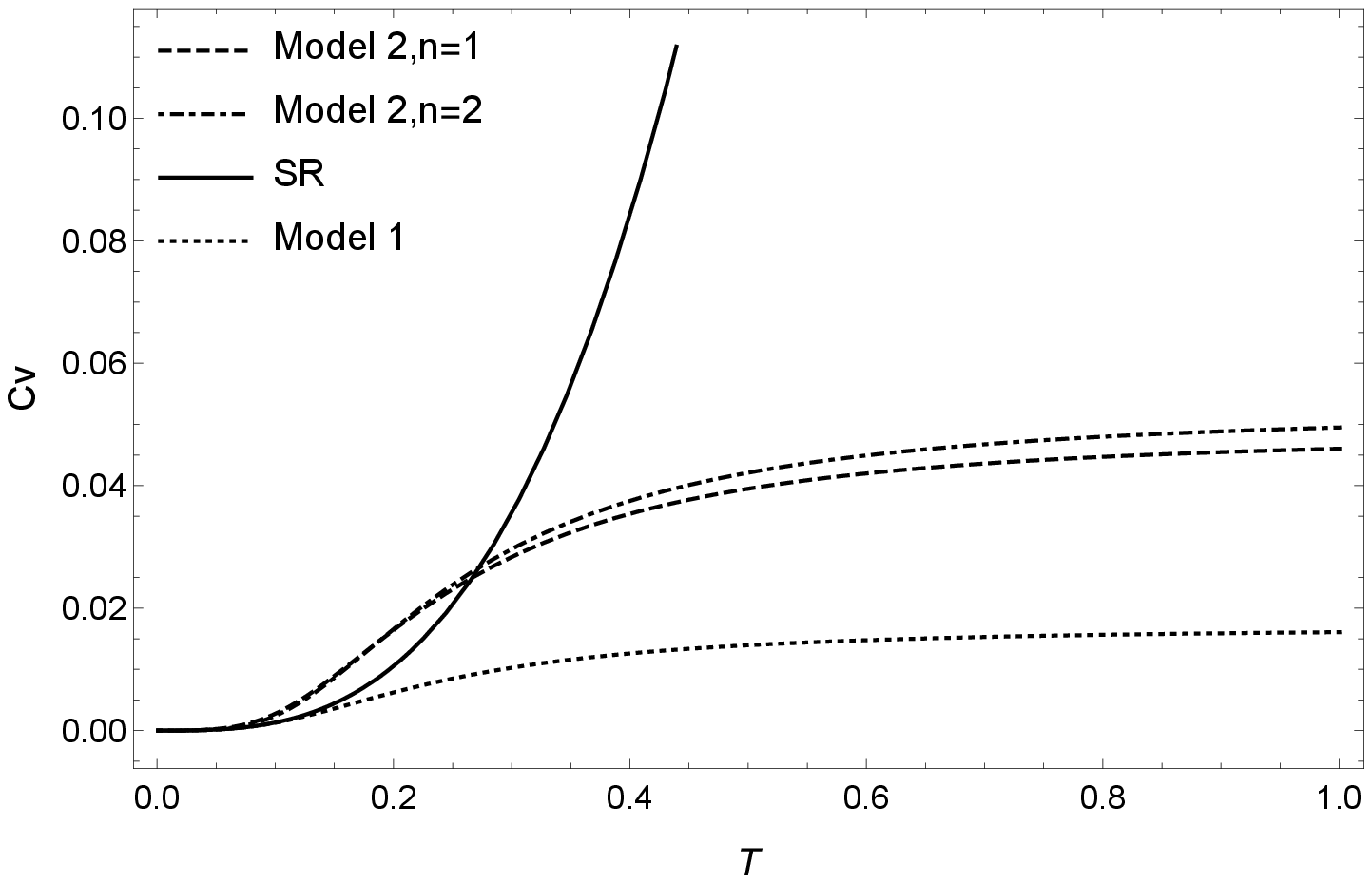}
  \caption{ Specific heat $C_v$, versus temperature }
  \label{fig:sub2}
\end{subfigure}
\caption{
	In figure 1 we have plotted, pressure $P$, Entropy $S$, specific heat $C_v$ of photon gas  against temperature
$T$ for model 1 (DSR), model 2 with $n=1$ and $n=2$.
The  results of model 1 and 2
have been obtained numerically using $Mathematica$\protect.
We
have used the Planck units, and the corresponding parameters
take the following values: $E_P=1$, $k_B = 1$, $V = 1$, $h = 1$, $c=1$.
In this scale,  $T = 1 $ is
the Planck temperature, $T_P$.
}
\label{fig:test}
\end{figure}
In this paper we have studied the thermodynamics of photon gas with two different models: Model 1 is based on DSR that preserves Lorentz algebra and Model 2 is  based on a  phenomenological Lorentz violating quantum gravity model. 
But, due to the presence of the deformed density of states and an upper bound, the quest
of solving the partition function
becomes complicated.
The results of our calculation
from both of the models 
coincide with the
known results \cite{path} of SR theory in the limit $E_P\rightarrow\infty$, as expected. 
In model 1,  we have seen that the  growth of all the thermodynamic quantities are slower compared to SR. This is due to the presence of an invariant energy
upper bound in this theory.  Therefore the microstates can attain
energies only up to a finite cutoff, whereas in the SR
theory, the microstates can attain energies up to infinity.
The current study provides significant improvement of the previous study based on model 1.
We have not only corrected the mathematical error of previous study\cite{mmfc} but have noticed that the final outcome is also changed  compared to ref. \cite{mmfc} due to the correction. The complete evaluation of equation (20), (21) and (22)
reveal in figure 1 that the thermodynamic quantities in this model change more slowly with temperature than previously reported.
Regarding the  second model under consideration,
it was reported by Camacho and Macias that the thermodynamic quantities of photon gas would have higher value 
in contrast  to special relativity in this model.
But as we have introduced an upper bound in this model, we have found the result to be quite the opposite. Though the thermodynamic quantities have greater values than those compared to SR initially, but at a certain temperature both graphs intersect and afterwards the thermodynamic quantities in this Lorentz violating model attain lower value compared to SR. This exact behaviour is also noticed a  recent study based on another Lorentz violating model\cite{mmf.kazi}. So these can be taken as signatures for future experiments to detect 
spacetime behaviour near Planck scale, whether it admits Lorentz violating or preserving structure. This property was not noticed
in the previous study based on model 2 due to the absence of upper energy bound. Therefore a significant improvement of statistical mechanics study is achieved  based on model 2.\\\\ 
As we have seen that, deformed dispersion relation and upper energy bound 
cause
the modification of the thermodynamical behavior
as well as the
equation of state of the massless bosons such as photon gas, we are most certain it will also change such characteristics for massive bosons.
We are planning to re-examine the 
models for massive bosons to find out the properties of Bose Einstein condensates near Planck scale. 
This  may have interesting effect in the scalar field dark matter model\cite{dm}, where
the dark matter particle is conjectured as a spin-0 boson.
In a quite similar spirit we may study the case of fermions in these models with upper energy models. This will yield an
intriguing situation in astrophysics, since the Chandrasekhar mass–radius relation for white
dwarfs is a direct consequence of the Fermi statistics. 
We are in a process to generalize standard cosmology results within these models.
\section*{Appendix: Evaluation of Integral}
The integrals that appear in this paper are of the form
\begin{equation}
    I_n=-\int_{0}^{E_p}E^{n-2}\ln\left(1-e^{- E/k_BT}\right)dE
\end{equation}
Note that we have put back Boltzmann constant $k_B$. Integrating by parts we have
\begin{equation}
    I_n=- \frac{E_p^{n-1}}{n-1}\ln(1-e^{-T_p/T})+\frac{(k_BT)^{-1}}{n-1}I
\end{equation}
Where 
\begin{equation}
    I=\int_{0}^{E_p}\frac{E^{n-1}e^{- E/k_BT}}{1-e^{-E/k_BT}}dE
    =\int_{0}^{E_p}\left(\sum_{m=1}^{\infty}E^{n-1}e^{-mE/k_BT}dE\right)
\end{equation}
and $T_p=E_p/k_B$ is the Planck temperature. In the scale $k_B=1$ which has been used throughout the paper we have $E_p=T_p$. The first term tends to zero because of the very large value of $E_p$. Substituting $t=m E/T$ and finally changing the order of sum and integral we obtain,
\begin{equation*} 
    I=(T)^{n}\sum_{m=1}^{\infty}\int_{0}^{mT_p/T}\frac{t^{n-1}e^{-t}}{m^n}dt\\   
\end{equation*}
Therefore
\begin{equation}
    I =(T)^{n}\sum_{m=1}^{\infty}\frac{\gamma(n,mT_p/T)}{m^n}
\end{equation}
where $\gamma(a,z)$ is the lower incomplete gamma function which is related to the gamma function by $\gamma(a,z)=\Gamma(a)-\Gamma(a,z)$ where $\Gamma(a,z)$ is the upper incomplete gamma function defined by  $\int_{z}^{\infty}x^{a-1}e^{-x}dx$. Finally we have
\begin{equation}
    I_n=\frac{(T)^{n-1}}{n-1}f_n
\end{equation}
Where 
\begin{equation}
    f_n=\sum_{m=1}^{\infty}\frac{\gamma(n,mT_p/T)}{m^n}
\end{equation}
In the limit $E_p\rightarrow\infty$, $f_n$ becomes $f_n=\zeta(n)\Gamma(n)$. 
Finally let us note the derivative of $f_n$ with respect to $T$
\begin{equation}
    f'_n=-\frac{T_p^{n}}{T^{n+1} \left( \exp(T_p/T)-1\right)}
\end{equation}
Note that both at $T\rightarrow 0$ and $T_p\rightarrow \infty$, $f'\rightarrow 0$.
\section*{Acknowledgement}
SB would like to thank Supantho Raxit for proofreading the manuscript.

\end{document}